\documentstyle[aps,twocolumn]{revtex}

\begin{document}

\noindent
{\bf Comment on ``Resonant Spectra and the Time Evolution of
the Survival and Nonescape Probabilities''}

\vspace{5mm}

Some time ago, Garc\'\i a-Calder\'on, Mateos and Moshinsky
\cite{Moshinsky} investigated the time evolution of the quantum
decay of a state initially located within an interaction region
of finite range. In particular, they showed that the survival 
$S(t)$ and nonescape $P(t)$ probabilities behave differently
at large times: $S(t)\sim t^{-3}$, whereas $P(t)\sim t^{-1}$
when $t\to\infty$. The purpose of this Comment is to show,
following their analysis, that $P(t)\sim t^{-3}$ also ---
a result already reported in the literature \cite{Moyses}.

According to Ref.\ \cite{Moshinsky}, $P(t)\sim t^{-1}$
asymptotically because (Eq.\ (21); equation numbering follows
Ref.\ \cite{Moshinsky})
\begin{displaymath}
\Delta\equiv\sum_{r=-\infty}^{\infty}\sum_{s=-\infty}^{\infty}
\frac{C_r^*C_sI_{rs}}{k_r^*k_s}\ne 0.
\end{displaymath}
However, $\Delta=0$, instead. To prove this, we use the
definition of $I_{rs}$ [Eq.\ (15)],
\begin{displaymath}
I_{rs}=\int_0^R u_r^*(r)u_s(r)\,dr,
\end{displaymath}
to rewrite $\Delta$ as
\begin{displaymath}
\int_0^R f^*(r)f(r)\,dr,\qquad f(r)\equiv\sum_{n=-\infty}^{\infty}
\frac{C_nu_n(r)}{k_n}.
\end{displaymath}
Using the definition of $C_n$ [Eq.\ (12)], the function $f(r)$
can be rewritten as
\begin{displaymath}
f(r)=\int_0^R \psi(r',0)\left(\sum_{n=-\infty}^{\infty}
\frac{u_n(r')u_n(r)}{k_n}\right)dr',
\end{displaymath}
where $\psi(r,0)$ is the wave function at $t=0$. Since the
term in parentheses vanishes [Eq.\ (7)], it follows that
$f(r)=0$, which completes the proof.

To prove that $P(t)\sim t^{-3}$, it is convenient to work 
directly with the time-dependent Green function,
$g(r,r';t)$, which, for $(r,r')<R$, can be written as [Eq.\ (9)]
\begin{displaymath}
g(r,r';t)=\sum_{n=-\infty}^{\infty}u_n(r)u_n(r')M(k_n,t).
\end{displaymath}
Using Eqs.\ (18) and (19), and the relations $k_{-n}=-k_n^*$,
$u_{-n}(r)=u_n^*(r)$, one can find the asymptotic expansion
of $g(r,r';t)$:
\begin{eqnarray*}
g(r,r';t)&\approx&\sum_{n=1}^{\infty}u_n(r)u_n(r')\,e^{-ik_n^2t} \\
&+&\frac{A}{t^{1/2}}\sum_{n=-\infty}^{\infty}\frac{u_n(r)u_n(r')}{k_n} \\
&+&\frac{B}{t^{3/2}}\sum_{n=-\infty}^{\infty}\frac{u_n(r)u_n(r')}{k_n^3}
+\ldots, 
\end{eqnarray*}
where $A$ and $B$ are constants. Again, because of Eq.\ (7), the 
coefficient of $At^{-1/2}$ cancels out exactly and, therefore,
$g(r,r';t)\sim t^{-3/2}$ when $t\to\infty$. Since
\begin{displaymath}
P(t)=\int_0^R\psi^*(r,t)\psi(r,t)\,dr,
\end{displaymath}
and
\begin{displaymath}
\psi(r,t)=\int_0^R g(r,r';t)\psi(r',0)\,dr',
\end{displaymath}
it follows that $P(t)\sim t^{-3}$ asymptotically.

I thank Carlos Alberto Arag\~ao de Carvalho for a critical reading
of this note. This work had financial support from CNPq.

\vspace{5mm}
\noindent
R.\ M.\ Cavalcanti \\ 
Departamento de F\'\i sica \\
Pontif\'\i cia Universidade Cat\'olica do Rio de Janeiro \\
Brasil

\vspace{5mm}

\end{document}